# Metal Coupled Folding of Cys$_2$His$_2$ Zinc-Finger

Wenfei Li, Jian Zhang, Jun Wang, and Wei Wang[*]

*National Laboratory of Solid State Microstructure, and Department of Physics,*

*Nanjing University, Nanjing 210093, China*

Email: wangwei@nju.edu.cn

*Abstract*: Zinc-fingers, which widely exist in eukaryotic cell and play crucial roles in life processes, depend on the binding of zinc ion for their proper folding. To computationally study the zinc coupled folding of the zinc-fingers, charge transfer and metal induced protonation/deprotonation effects have to be considered. Here, by attempting to implicitly account for such effects in classical molecular dynamics and performing intensive simulations with explicit solvent for the peptides with and without zinc binding, we investigate the folding of the Cys$_2$His$_2$ type zinc-finger motif and the coupling between the peptide folding and zinc binding. We find that zinc ion not only stabilizes the native structure, but also participates in the whole folding process. It binds to the peptide at early stage of folding, and directs or modulates the folding and stabilizations of the component β-hairpin and α-helix. Such a crucial role of zinc binding is mediated by the packing of the conserved hydrophobic residues. We also find that the packing of the hydrophobic residues and the coordination of the native ligands are coupled. Meanwhile, the processes of zinc binding, mis-ligation, ligand exchange and zinc induced secondary structure conversion, as well as the water behaviour, due to the involvement of zinc ion are characterized. Our results are in good agreement with related experimental observations, and provide significant insight into the general mechanisms of the metal-cofactor dependent protein folding and other metal induced conformational changes of biological importance.

# Introduction

Zinc-fingers represent a wide class of proteins for which the folding depends on the binding of metal-cofactors. They play crucial roles in regulating the gene expression and other biological processes by interacting with DNA, RNA and proteins[1-4]. Disruption of their native structures may lead to severe human diseases, e.g., cancer and neurological disorders[3]. Understanding the folding and stability of the zinc-fingers is fundamental for treating the related diseases, and is also helpful for developing better strategies for the zinc-finger based *de novo* protein design[5-9]. In particular, because the processes, e.g., metal binding, ligand exchange and metal induced secondary structure conversion exhibited by the zinc-finger due to the involvement of zinc ion (Zn(II)), may also occur during the folding of other metalloproteins, knowledge of the metal coupled folding for this motif can provide insight into the general mechanisms of the metal-cofactor dependent protein folding, which is a fundamental but not well understood problem[10-13]. A lot of experimental work is devoted to identifying the factors which affect the folding and stabilization of the zinc-fingers, and many features have been revealed[4,14-25]. However, a detailed atomistic picture of how the Zn(II) coordinates to its native ligands and how such coordination process couples with conformational motions is still lacking because of the limited temporal and spatial resolutions in experiments. It is even unclear whether the Zn(II) binds to the peptide at early stage of folding and induces the peptide folding, or the Zn(II) just binds to the peptide at the end of folding and stabilizes the native structure[26,27].

Molecular dynamics (MD) simulation is a powerful tool for revealing the detailed mechanism of protein folding with the complements of experimental data due to its atomic resolution[28-32]. However, the involvement of metal ion in the zinc-fingers makes the simulation much complicated and no simulation work has been reported. The reason for this is that the formation and breaking up of the coordination bonds between Zn(II) and its ligands occur frequently and charge can transfer between the Zn(II) and the liganding atoms during the folding/unfolding of the peptide (Figure 1c). To reasonably describe such effect, quantum mechanics should be employed. However, such quantum mechanics process is difficult to be included into the simulations of peptide folding because of the tremendous computational demands. Additionally, the titratable ligands can be deprotonated with the formation of coordination bonds due to strong oxidative activity of Zn(II), which contributes to the experimentally observed enthalpy change significantly[23] (Figure 1d). The proton transfer during this zinc induced protonation/deprotonation process can result in the electrostatic field redistribution and affect dramatically the folding kinetics. Without considering these effects, it is impossible to properly characterize the metal coupled folding properties theoretically. These difficulties hindered the applications of the classical MD in studying the metal coupled folding of zinc-fingers and other metalloproteins.

In literature, several models were proposed for describing the coordination bonds between Zn(II) and liganding atoms[35-38]. Among them, the non-bonded model presented by Stote and Karplus is widely used in studying the structure and dynamics of zinc containing proteins around their native states[35]. In this model, the coordination bonds are treated by the electrostatic and van der Waals (vdW) potentials. Based on this model, Sakharov and Lim considered the effects of the charge transfer between Zn(II) and native ligands and the local polarization around the Zn(II)[34]. With this model, the coordination number and geometry in the native state of the $Cys_2His_2$ and $Cys_4$ type zinc-fingers can be well described. In fact, the importance of the charge transfer and polarization effect were addressed more than ten years ago by Gresh and coworkers, and the SIBFA polarizable force field was developed to include these effects[39-42]. Recently, this force field was applied to calculate the relative stabilities of a number of conformations for a $Cys_2HisCys$ type zinc-finger with 18 residues, and the NMR structure is successfully predicted to be the global minimum[43].

However, to simulate the zinc coupled folding of the zinc-fingers, the zinc induced protonation/deprotonation effect and the charge transfer between the Zn(II) and all the potential ligands have to be included in the simulations as discussed above. In this work we report a first simulation study on the zinc coupled folding for the second finger of the human transcription factor Sp1 (Sp1f2)[44] which is a typical Cys$_2$His$_2$ type zinc-finger (Figure 1a,b) by implicitly accounting for the effects of the charge transfers that occur between Zn(II) and its ligands as well as for protonation/deprotonation effects in the classical MD model and performing intensive all-atom simulations for the peptides with and without zinc binding. To our knowledge, this is the first simulation study on the coupling between the metal binding and peptide folding.

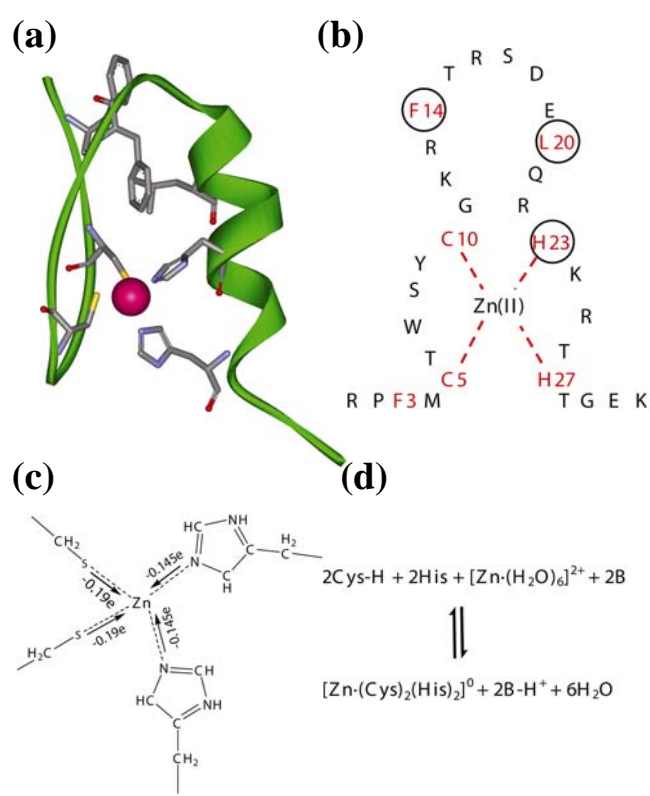

*Figure 1.* Structure, sequence, coordination bonds and protonation process of Cys$_2$His$_2$ zinc-finger. (a) The native structure of Sp1f2 which contains 31 residues with Phe3, Phe14 and Leu20 being the conserved hydrophobic residues and Cys5, Cys10, His23 and His27 being the conserved ligand residues. (b) The sequence of Sp1f2. The conserved residues are represented with red color, and the residues of the hydrophobic core are circled. (c) The coordination bonds formed between the Zn(II) and Cys5, Cys10, His23 and His27. The numbers along the arrows are the transferred charge calculated using Gaussian 98 based on model systems[33,34]. (d) The zinc induced protonation/deprotonation process with the formation of the coordination bonds between the Zn(II) and the cysteines.

# Methods

**1. Coordination bonds.** The simulations are performed using AMBER8 package with water molecules being treated explicitly[45]. The AMBER ff03 force field is used with some modifications[46,47]. The bonds formed between Zn(II) and potential liganding atoms are described by the non-bonded model[35]

$$V_{Zn-i} = \frac{q_{Zn}q_i}{4\pi\varepsilon_0 r_{Zn-i}} + 4\varepsilon_{Zn-i}\left[\left(\frac{\sigma_{Zn-i}}{r_{Zn-i}}\right)^{12} - \left(\frac{\sigma_{Zn-i}}{r_{Zn-i}}\right)^{6}\right]. \quad (1)$$

Here $q_{Zn}$ and $q_i$ are the charges of the Zn(II) and the $i$-th liganding atom, $r_{Zn-i}$ is the distance between them; and $\varepsilon_{Zn-i}$ and $\sigma_{Zn-i}$ are the combined vdW parameters. The vdW parameters of the Zn(II) are set as $\varepsilon_{Zn} = 0.183$ kcal/mol and $\sigma_{Zn} = 1.57$Å which can reproduce the experimental hydration free energy of Zn(II) relative to the experimental hydration free energies of 22 other dications[34,48]. In this work, a transfer of charge is considered to occur for the $S_\gamma$ of the cysteines, $N_\varepsilon$ of the histidines, O and N of the backbone, O of the side chains in the Glu, Asp, Gln, Ser, Thr and Tyr, and the O in water molecules. The charge transferred from one of the ligand atoms to the Zn(II) at any distance is obtained by $\Delta q_i(r_{Zn-i}) = A_i r_{Zn-i} + B_i$ with the coefficients $A_i$ and $B_i$ being extracted based on the quantum chemical method and the AMBER ff03 parameters[46,47] (See Ref. 34, and Table S1 for more details for the calculations of the coefficients $A_i$ and $B_i$.). Thus the atomic charges of all the ligand atoms and Zn(II) at every time step can be calculated based on the atomic charge of AMBER ff03 force field. The model systems used to mimic the ligands in deriving the parameters based on quantum chemical calculations and the derived parameters $A_i$ and $B_i$ are presented in Table S1. The systems are optimized at B3-LYP/6-31+G* level as done in the previous work by Sakharov and Lim[34].

As the electrons in the imidazole ring are highly delocalized, binding of the Zn(II) can result in dramatic polarization of the whole imidazole ring (charges are redistributed among the atoms of the imidazole ring)[49]. This effect is not included in the original AMBER ff03 force field. Therefore, in this work, the charges of the imidazole ring are adjusted by a distance dependent manner to describe this effect. When the distance between the Zn(II) and the $N_\varepsilon$ of imidazole ring comes close, the charge of the liganding $N_\varepsilon$ is increased with the charges of the rest atoms in the imidazole ring being scaled accordingly. Since the total charge of the imidazole ring for the neutral histidine is close to zero, such adjustment results in a dipole in the imidazole ring, which can mimic the Zn(II) induced polarization of the imidazole ring. The increment of the charge for the $N_\varepsilon$ is determined by fitting the experimental coordination geometry. The finally obtained charge of $N_\varepsilon$ is -0.70e. The charges of the imidazole ring are restored to the values of the original AMBER ff03 force field linearly with the increasing of the Zn(II)-$N_\varepsilon$ distance, namely, the polarization of the imidazole ring vanishes when the Zn(II)-$N_\varepsilon$ distance is larger than a threshold (In this work, this distance threshold is set as two times of the sum of their vdW radii). During this process, the total charge of the histidine is not changed. As the electrons in the side-chain of cysteine are not as delocalized as in the imidazole ring, the polarization effect of the cysteine is not explicitly treated, and only the charge transfer effect is included.

Although such an implicit treatment can only partly include the induced polarization effect, the experimental coordination geometry of the zinc-finger can be well reproduced (Figure S1 and Table S2). We compared the calculated coordination geometry of the classical zinc-finger with those calculated by the model of Sakharov and Lim [34], and by the AMBER ff02 polarizable fore field[50] with charge transfer. We found that the present model is comparable with the ff02 polarizable force field with the similar treatment of charge transfer and the model by Sakharov and Lim in reproducing the experimental coordination geometry of classical zinc-finer peptide (Table S2). Noting that the length of the coordination bond should be combined with other quantities, such as the coordination number and *rmsd*, to be used as benchmarks for validating the model because of the limited resolutions of most crystal structures.

**2. Protonation effects.** In the realistic protonation process, protons used to protonate the titratable groups are taken from solvent pool. At the same time, buffer supplies protons to the solvent pool for maintaining constant pH value. To fully describe this process, quantum chemical calculations should be employed. However, such calculation is limited to the simulations of short time scale because of its tremendous computational cost. Meanwhile, explicitly including the solvent protons in simulation box will result in very low pH value. To avoid such problems, in the present simulations the protons needed to protonate the residues are supplied directly by buffer ions with the protons being considered as a unit of positive charge. In this way, the protonation process is described as the transferring of a unit of positive charge from buffer to the titratable residues. Note that one can calculate the protonation state and its pH dependence using the constant pH MD[51], in which the free energy differences between each protonation state need to be calculated. In the present simulations the protonation state of the ligands before and after the formation of the bonds are predefined according to the experimental data, which makes the simulations much simpler, and the calculation of free energy differences can be avoided. As the deprotonation of the titratable residues is resulted from the zinc binding, the transferred charges between the buffer and the titratable residues are dependent on the distance between the ligand and the Zn(II). In this work, a linear dependence is used. At the equilibrium distances between the Zn(II) and the titratable residues, both the cysteines and histidines adopt deprotonated state (the charge states of the deprotonated cysteine and histidine in the AMBER ff03 force field are used). With the breaking up of the coordination bonds, the cysteines are assigned as protonated form gradually (the charge states of the protonated cysteine in the AMBER ff03 force field is used except that the charge of the $H_\gamma$ is merged into the $S_\gamma$ since the $H_\gamma$ is not explicitly represented in the simulation). Here, the positive charge needed to protonate the cysteine is supplied by the buffer which is represented by dissociated NaCl as discussed below, namely, the negative charge of the cysteine is progressively lost and transferred to the "buffer" ions during the protonation. These "buffer" ions are solvated in the water box, and their motions are coupled with the water bath. Because the histidines have certain probability to be protonated in the unbound peptide at neutral pH value[52], there should be protons being transferred from the buffer to the histidines with the breaking up of the Zn(II)-His bonds. Here, the charge of 0.3e is transferred from the buffer to the histidine to describe this process since around 0.3 proton is needed averagely to protonate each histidine with the breaking of the corresponding bond according to its pKa value[23,52]. Therefore, upon the breaking up of the four native coordination bonds, the net charge of the complex (including peptide and Zn(II)) increases by +2.6e (from +5e to +7.6e) due to the full protonation of the cysteines and the partial protonation of the histidines. Noting that the final charges of the liganding atoms are the total effects of the charge transfer, polarization and the Zn(II) induced protonation/deprotonation effects.

In our simulations, the buffer is not represented realistically. Instead, we use NaCl to mimic the buffer, and the charges needed to protonate the cysteines and histidines are supplied by the Na$^+$. Four NaCl are added to model the buffer ions. This corresponds to a concentration of 56 mM which is within the range of typical buffer concentration used experimentally. Though the protons are not represented explicitly and the proton transfer is treated as charge transfer between the buffer and the titratable residues, the resulted variation of the electrostatic field due to the proton transfer, which has significant contributions to the conformational distribution of the peptide, can be basically characterized.

**3. Simulation** In the simulations, the nuclear magnetic resonance (NMR) structure of the Sp1f2 (PDB code: 1sp2)[44] is solvated in a TIP3P water box. The replica-exchange MD (REMD) method is used for conformational sampling[53,54]. Totally 50 ns are simulated for each of the 64 replicas with the temperatures ranging from 289 to 607K. The convergence of the simulation is monitored, and presented in Figure S2, Figure S3 and Figure S4. The structures of the last 40 ns for each replica are used for analysis. In constructing the free energy landscape at certain temperature, the weighted histogram analysis method (WHAM) is used[55,56]. In the WHAM, the density of state for the system is given by

$$W(E,R) = \frac{\sum_n N_n(E,R)}{\sum_n n_n \exp(f_n - \beta_n E)},$$ where the $N_n(E,R)$ is the energy histogram with reaction coordinate $R$

and potential energy $E$ for the conformations sampled at the $n$th temperature $T_n$, and $\beta_n = k_B T_n$ with $k_B$ being the Boltzmann constant. The $n_n$ in the denominator is the number of structures sampled at $T_n$. The $f_n$ in the density of state is determined by $\exp(-f_n) = \sum_{E,Q} W(E,R)\exp(-\beta_n E)$. The density of state $W(E,R)$ can be obtained consistently by iterating these two formulae. With the density of state, we can obtain the partition function and the free energy landscape projected onto the reaction coordinate $R$. Here, $R$ refers to any reaction coordinates. By using the WHAM, the partition function at certain temperature can be determined more precisely by combining the structures sampled at different temperatures.

For comparison, we also conducted a control simulation with the Zn(II) being removed and the cysteines being protonated. The reaction coordinates $Q$, $Q_\beta$, $N_\alpha$, $R_g$, $R_g^{core}$, $rmsd$ and $N_{nl}$ are used in this work. $Q$ is the fractional native contact. $Q_\beta$ is the fractional native contacts for residues Arg1-Arg16. $N_\alpha$ is the number of helical residues formed among residues Ser17-Lys30. $R_g$ and $R_g^{core}$ are radius of gyration of the peptide and the hydrophobic core, respectively. $rmsd$ is the rms distance for all atoms. $N_{nl}$ represents the number of the native ligands coordinated.

## Results and discussions

**1. Binding order of the native ligands.** In order to reveal the metal coupled folding mechanism, we first study the binding order of the four native ligands, namely, Cys5, Cys10, His23 and His27 to the Zn(II), which is obtained by performing correlation analysis to the sampled structures (Figure 2a). The percentages for the formation of the four native coordination bonds are calculated under the condition of one, two, three and four native ligands formed, respectively. We consider the coordination bonds formed if the distance between the Zn(II) and the liganding atoms is less than 2.5 Å.

From Figure 2a we can see that the bonds Zn(II)-Cys5 and Zn(II)-Cys10 are formed earlier than the Zn(II)-His23 and Zn(II)-His27. Either the Cys5 or Cys10 could be the first native ligand coordinated with a percentage of 75% for the Cys10 and 25% for the Cys5. Among the two histidines, the His23 coordinates to the Zn(II) prior to the His27. This binding order is consistent with the experimental observation based on Raman spectra[22] and Uv-vis absorption spectra[17], indicating that our model can describe the zinc binding process reasonably. Such stepwise binding of the native ligands to the Zn(II) results in formations of component secondary structures and tertiary structure (Figure 2b), which will be discussed later in this paper. Note that before all the four native bonds are formed, other non-native ligands, i.e., water molecules or/and atoms from other residues, can be coordinated to the Zn(II) besides the native ligands, which makes the coordination number saturated.

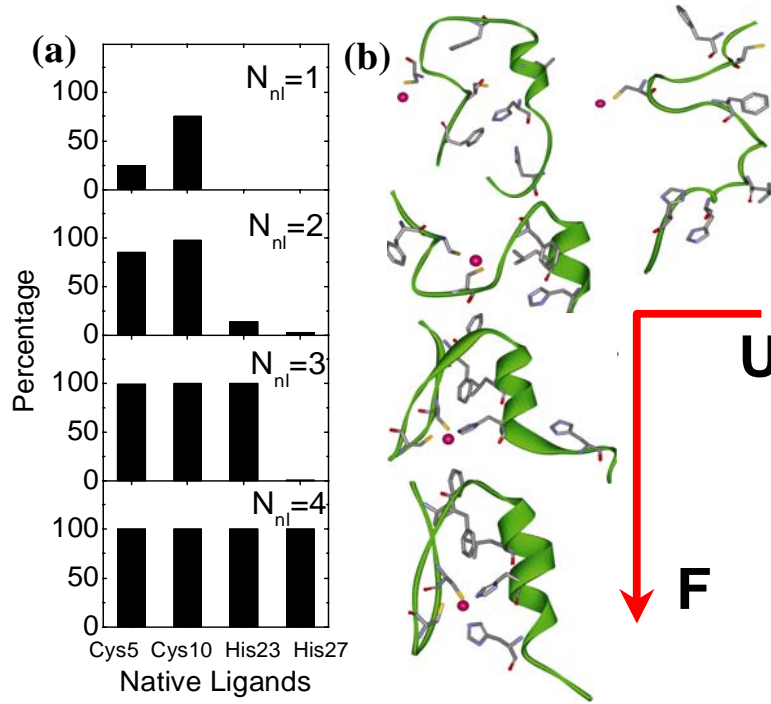

*Figure 2.* (a) Percentages for the formation of the four native coordination bonds with one, two, three, and four native coordination bonds formed. $N_{nl}$ is the number of the native coordination bonds formed. (b) Schematic representation of the effects of native coordination bond formation on peptide folding.

**2. Ligand exchange**. Figure 3a shows the ligand exchange process during the peptide folding at melting temperature $T_m$ which is obtained by fitting the thermal denaturation curves (See Figure S5 for the details of the fitting and derivation of the $T_m$). Here, the native ligand number is defined as the averaged number of the native ligands, namely, Cys5, Cys10, His23 and His27 coordinated to the Zn(II) at $Q$. Similarly, the non-native ligand number is defined as the averaged number of the non-native ligands, namely, the protein groups except for the native ligands, coordinated to the Zn(II) at $Q$. The total coordination number is the number of all ligands coordinated to the Zn(II) at $Q$. A cutoff of 2.5 Å is used to define the coordination bond. One can see that around four water molecules and one native ligand are coordinated to the Zn(II) at unfolded state. With the folding of the peptide, the water molecules are gradually expelled from the first ligand shell. At native state (Q>0.9), all the native ligands are coordinated, and the water molecules come out of the ligand shell. It is interesting to note that other protein atoms can also come into the ligand shell before the native state is arrived (red line with solid squares). Figure 3b shows the percentage for each protein atom of the peptide to coordinate with the Zn(II). The four atoms with the highest coordination probability are the natively coordinated atoms. One can see that besides the native liganding atoms, other atoms around and between two cysteines and at the C-terminal have higher probability to coordinate with the Zn(II). Such mis-ligation mainly occurs at the later stage of the folding as suggested by the broad peak around 0.65<Q<0.85 in Figure 3a. In fact, these non-native bonds mainly exist in the conformations with two or three native ligands coordinated (Figure S6). For these conformations, the last step of folding involves the replacement of the mis-coordinated ligands and water molecules by the native ligands. It is worth pointing out that such mis-ligation and ligand exchange were proposed recently in a number of experiments to explain the observed folding kinetics of other metalloproteins[10]. Our simulation work provides direct observations of such processes theoretically, and confirms the mechanism of mis-ligation and ligand exchange.

In order to provide a more detailed picture for the effects of coordination bond formations, we investigated the variation of the averaged protonation states for the four native ligands along the folding pathway of the zinc-finger peptide. The results are presented in Figure S7. From Figure S7a, one can see that the Cys5 is mostly protonated when the $Q$ is small. In comparison, the Cys10 is only protonated by around 20%. This is because that the Zn(II)-Cys5 bond is mostly broken up at unfolded state, while the Zn(II)-Cys10 bond is still formed to large extent as shown in Figure S7b. At unfolded state, both the His23 and His27 are protonated by 30% because the Zn(II)-His23 and Zn(II)-His27 are all broken up.

With the increasing of the reaction coordinate $Q$, these ligands are gradually deprotonated due to the coordinations with the Zn(II).

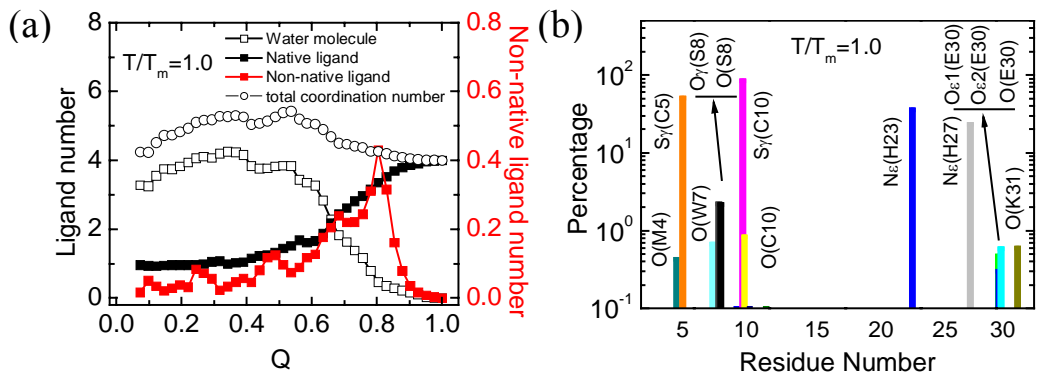

*Figure 3.* Number of coordinated native ligands, water molecules and other non-native ligands, as well as the total coordination number as a function of $Q$ (a) and the percentage of structures with each protein atom coordinating with the Zn(II) (b). Noting that the non-native ligand number in (a) is measured by the scale of the right axis.

**3. Folding pathway of zinc-finger peptide**. The free energy landscape is very useful in elucidating the folding properties of proteins[57-59]. We characterize the folding mechanism of the zinc-finger with the help of Zn(II) based on the free energy landscape projected onto reaction coordinates ($R_g$, $Q$), ($rmsd$, $Q$), ($Q_\beta$, $Q$) and ($N_\alpha$, $Q$) (see **Methods** for their definitions) as shown in Figure 4a-d. For comparison, the free energy landscape for the peptide without zinc binding is also constructed (inset of Figure 4a, and Figure S8). These landscapes are calculated at melting temperature $T_m$ since at this temperature both the unfolded states and the folded state can be sampled more sufficiently. In Figure 4a three major basins corresponding to two intermediates I1 and I2, and the native state N are mostly populated besides the extended states. These major states can also be identified in other reaction coordinates. At the intermediate I1, the β-hairpin is basically unstructured as $Q_\beta \approx 0$ (Figure 4c). Part of the α-helix can be formed, but is unstable as indicated by the large variation of the $N_\alpha$ between 0 and 8 (Figure 4d). In comparison, at the intermediate I2, the β-hairpin is fully formed as $Q_\beta \approx 1$, and the α-helix is partially folded and stabilized since the conformations are mainly populated around $N_\alpha \approx 7$.

From Figure 4, a folding pathway of the peptide can be deduced. The folding initiates with a hydrophobic collapse, during which the α-helix may be partially formed although its stability is very low. Then the folding proceeds with full formation of the β-hairpin and partial formation and stabilization of the α-helix, which is a rate-limiting step since a high barrier needs to be overcome (Figure 4a-d). Finally, the folding finishes with full formation of the α-helix.

Note that such a folding pathway is based on the progression along reaction coordinates instead of time because the time information is mostly lost due to the use of the replica-exchange MD method[53,54]. To obtain the folding pathway corresponding to the time sequence, long time-scale simulations with the standard MD are needed, which is still a challenge to the present computational ability.

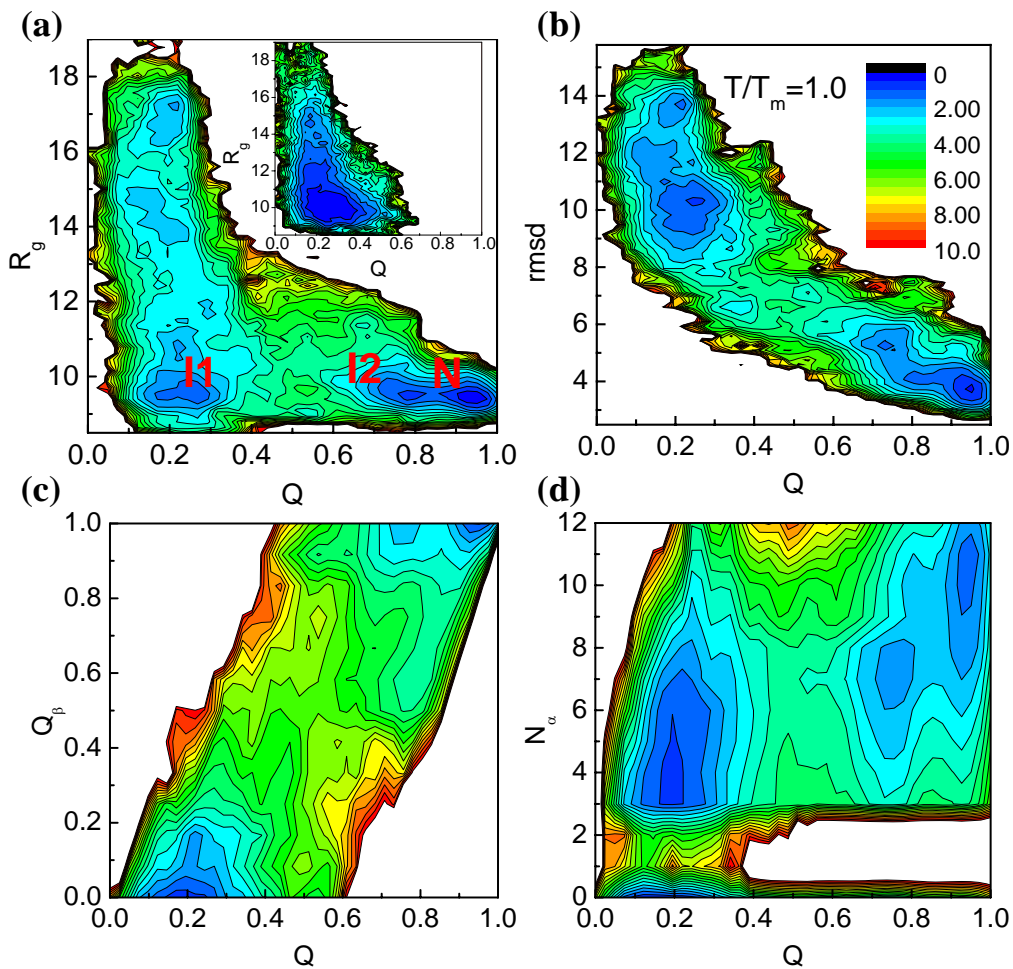

*Figure 4.* Free energy landscapes of peptide folding. The free energy projected onto reaction coordinates ($R_g$, $Q$) (a), ($rmsd$, $Q$) (b), ($Q_\beta$, $Q$) (c) and ($N_\alpha$, $Q$) (d) at $T_m$. The unit of the free energy is $k_B T_m$. The basin around the region of $Q \approx 0.95$ and $R_g \approx 9.4$ Å at (a), labelled as N, corresponds to the native state. While other two basins around the region of $Q \approx 0.3$ and $R_g \approx 9.6$ Å and the region of $Q \approx 0.8$ and $R_g \approx 9.4$ Å, labelled as I1 and I2, are related to two intermediates, respectively. Note that the barrier between 0 and 3 along the $N_\alpha$ axis in the left bottom of (d) results from the definition of the helical residues, which demands three or more consecutive residues satisfying the dihedral constraints (**Simulation details** in supporting information). For comparison, the free energy landscape for the peptide without zinc binding is presented in the inset of (a).

**4. Role of zinc binding on the zinc-finger folding.** The above folding scenario is strongly coupled with the binding of the Zn(II). To understand the role of the zinc binding on the folding of the zinc-finger peptide, we compare the conformational distributions of the peptide with (holo-peptide) and without (apo-peptide) zinc binding by projecting the free energy landscape onto several reaction coordinates (Inset of Figure 4a; Figure S8, Figure S9 and Figure S10). We can see that the Zn(II) has dramatic contributions to the folding free energy landscape. Instead of just changing the relative populations of each major state in the conformational space, binding of the Zn(II) completely changes the pattern of the free energy landscape by eliminating some major states and adding some new states. For example, when the Zn(II) is absent, both the native state N and the intermediate I2 disappear (Inset of Figure 4a and Figure S8). Meanwhile, the region $0.4<Q<0.5$ samples more conformations for the apo-peptide, while this region corresponds to the free energy barrier for the holo-peptide (Figure 4). Similar results can also be obtained from Figure S9 and Figure S10 which show the free energy landscapes of these two peptides projected onto the two most important principal components (PC1 and PC2) of the holo-peptide (Figure S9) and apo-peptide (Figure S10), respectively. Here, the two most important principal components correspond to the principle components with the largest eigen values obtained by diagonalizing the covariance matrix. One can see that the free energy landscapes in the principal component space are greatly different for the apo-peptide and holo-peptide. By performing cluster analysis to the structures of the holo-peptide and apo-peptide sampled at $T_m$, we find that the conformational distribution of the apo-peptide is more uniform compared to that of the holo-peptide (Figure S11), indicating that the structures of the apo-peptide are more heterogeneous than that of the holo-peptide. Meanwhile, the structures of the five most probable clusters are completely different for these two peptides (Figure S12). For example, the structure of the most probable cluster for the holo-peptide is similar to the structure of the native state of the zinc-finger. While the structure of the most probable cluster for the apo-peptide is relatively unstructured with some helical structures presented.

The above results indicate that the zinc binding not only stabilizes the native state, but also participates in the whole folding process. Without the help of the Zn(II), the zinc-finger peptide cannot fold to its proper three dimensional structure.

To reveal the molecular picture of how the zinc binding couples with the peptide folding, the free energy landscape is projected onto reaction coordinates ($N_{nl}$, $Q$) (Figure 5a). Three major basins can be seen in Figure 5a. The centers of these basins correspond to the states with one, three and four native coordination bonds formed, labelled as C1, C3 and C4, respectively. Before the coordination of the first native ligand, the β-hairpin is mostly unstructured, while part of the α-helix can form to certain extent (Figure 5b,c) due to the high intrinsic α-helix propensity according to the statistical data of Chou and Fasman[60] and the secondary structure prediction method APSSP2[61] (Figure S13 and Table S3). This indicates that the apo-peptide of zinc-finger is not totally unstructured, which is consistent with the result of Figure S12. The existence of α-helix structure in the apo-peptide can be supported by Raman spectra data for another $Cys_2His_2$ zinc-finger corresponding to the third finger of the mouse Zif268 protein in which the peak corresponding to α-helix at amide III region was observed clearly although the authors claimed that the peptide adopts β-strand structure without Zn(II)[22]. When the Zn(II) is coordinated with the first cysteine (C1), i.e., $N_{nl}$ = 1, the conformation of the peptide does not change significantly (Figure 5b,c). As the second cysteine is coordinated, a loop between the two cysteines is formed, which is helpful for the folding of the β-hairpin by decreasing the conformational space to be sampled. At this stage, the β-hairpin may occasionally form, and the α-helix content, particularly for the first two helical turns, increases (Figure S14a,b,c and Figure 5b,c). Note that this coordination state is unstable, and not dominantly populated as shown in Figure 5a. To emphasize its statistic insignificancy, we represent the percentages for $N_{nl}$=2 in Figure 5b and Figure 5c with dotted line and open symbols instead of solid line and closed symbols as used in other cases. When the His23 is also coordinated to the Zn(II) (C3), both the β-hairpin and the first two helical turns of the α-helix are well formed and stabilized (Figure 5b,c and Figure S14). From state C1 to C3 a high energy barrier is overcome, indicating that the coordination of the His23 to the Zn(II) is the most critical step for the peptide folding. Actually, once the His23 is coordinated, most of the secondary and tertiary structures are formed except for the last helical turn of the α-helix which is formed after the His27 is coordinated due to the geometrical constraint imposed by the coordinations of the His23 and His27 (C4) (Figure 5b). Such a crucial role of the Zn(II)-His23 coordination revealed above is compatible with an experimental observation that the structure of the peptide is severely disrupted when the His23 is mutated to Gly or Ala[21].

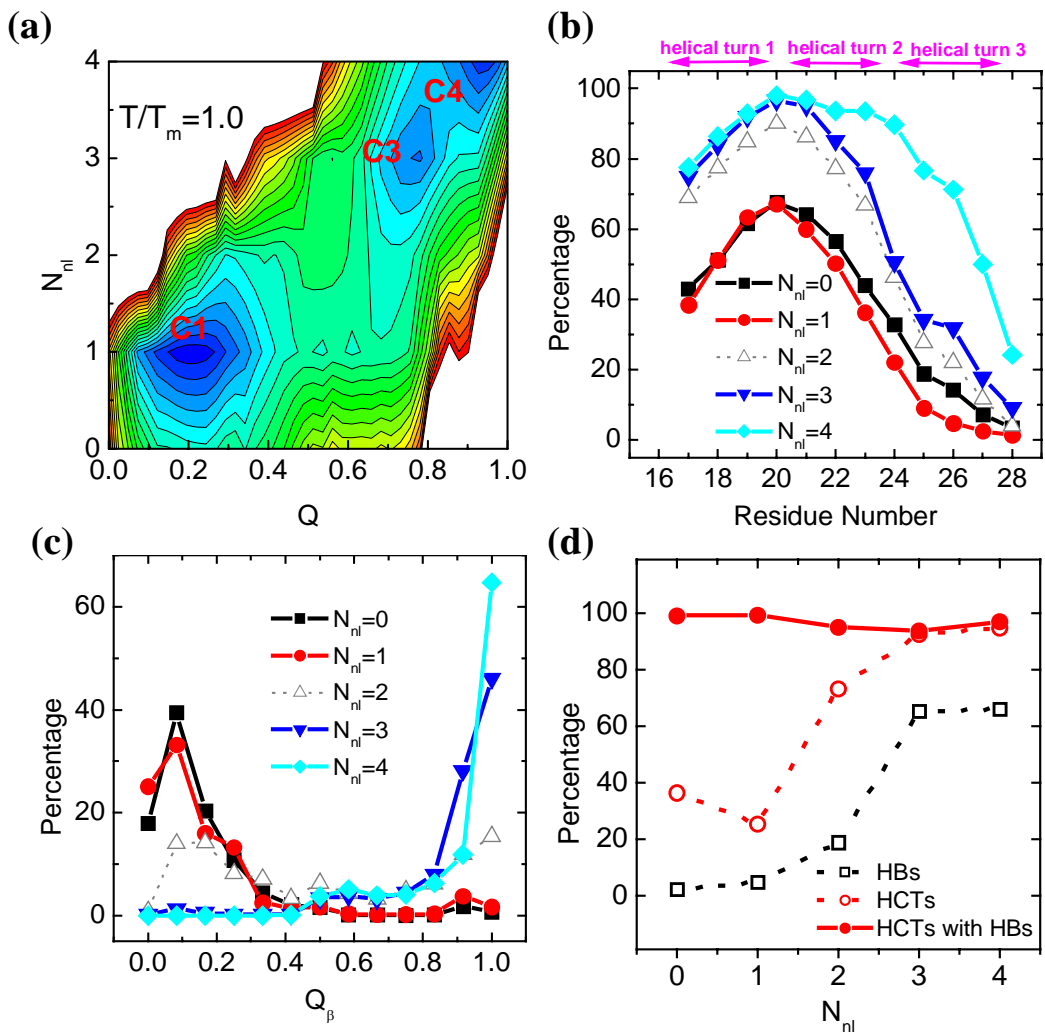

*Figure 5.* Role of the zinc binding on the peptide folding. (a) The Free energy landscape projected onto reaction coordinates ($N_{nl}$, $Q$) at $T_m$. The color scale is the same as that used in **Figure 4**. (b) Percentage of each residue adopting α-helix conformation with zero (black), one (red), two (gray), three (blue) and four (cyan) native ligands coordinated at $T_m$. (c) Distribution of the $Q_\beta$ for the conformations with zero (black), one (red), two (gray), three (blue) and four (cyan) native ligands coordinated at $T_m$. (d) Formation percentages for the two HBs in the β-hairpin (open squares) and the hydrophobic contacts (HCTs) between the side chain of the Phe3 and the residues of loop ranging from Phe14 to Leu20 (open circles) as a function of $N_{nl}$ at $T_m$. Meanwhile, formation percentage of the hydrophobic contacts under condition of the HBs formed (solid circles) is also shown.

The above results, together with the results of Figure 3, clearly demonstrate that the Zn(II) is involved in the whole folding process, and binding of it can direct and modulate the folding and stabilization of the component β-hairpin and α-helix.

**5. Cooperativity between formations of the hydrophobic core and the coordination bonds.** From Figure S14a, we can see that the packing and stabilization of this hydrophobic core need the coordinations of the Cys5, Cys10 and His23 to the Zn(II). Meanwhile, the packing of the hydrophobic core in turn stabilizes the above bonds. Such a cooperative stabilization between the hydrophobic core and the coordination bonds can be demonstrated in Table 1 which shows the percentage for the formation of the hydrophobic core with and without the formation of the coordination bonds, and the percentage for the formation of the coordination bonds with and without the formation of the hydrophobic core. We define the coordination bonds being formed when three or more native ligands are coordinated, and the hydrophobic core being formed when the radius of gyration of the hydrophobic core is less than 4.5 Å. One can see that the percentage for the formation of the hydrophobic core (coordination bonds) with the coordination bonds (hydrophobic core) formed is significantly higher than that without the formation of the coordination bonds (hydrophobic core), indicating that the formations of the coordination bonds and hydrophobic core is highly cooperative. This result clarifies the argument of whether the component structures of proteins stabilize the coordination bonds, or the coordination bonds stabilize the component structures of proteins during the folding of metalloproteins. The present results indicate that the component structures of proteins and the coordination bonds stabilize each other in a cooperative manner. The breaking up of either of them can destabilize the other structure.

*Table 1.* Percentage for the formation of the hydrophobic core with ($P(HC|CB)$) and without ($P(HC|\overline{CB})$) the formation of the coordination bonds, and the percentage for the formation of the coordination bonds with ($P(CB|HC)$) and without ($P(CB|\overline{HC})$) the formation of the hydrophobic core.

| $P(HC|CB)$ | $P(CB|HC)$ | $P(HC|\overline{CB})$ | $P(CB|\overline{HC})$ |
|---|---|---|---|
| 98.5% | 77.7% | 14.6% | 2.5% |

**6. Role of hydrophobic packing on the zinc-finger folding.** The above results show that although the His23 is in the C-terminal α-helix, the folding and stabilization of the N-terminal β-hairpin depend strongly on the Zn(II)-His23 coordination. Similarly, the folding and stabilization of the C-terminal α-helix also depend on the formations of the Zn(II)-Cys5 and Zn(II)-Cys10 which locate at the N-terminal β-hairpin. We will show that such a long-range correlation is mediated by the formation of the hydrophobic core consisting of Phe14, Leu20 and His23.

In the native structure of Sp1f2, two inter-strand hydrogen bonds (HBs) between Phe3 (CO) and Leu14 (NH) and between Cys5 (NH) and Lys12 (CO) are formed[44]. The β-hairpin is well folded only when these two HBs are formed. Therefore, the formation of the two HBs can be considered as a symbol of the β-hairpin folding. Previously, in the study of β-hairpin folding, it was found that contacts formed between hydrophobic residues from two β-strands are crucial for the formation and stabilization of the inter-strand HBs[62-66]. Similarly, for Sp1f2, we will show that the formation of two HBs also strongly depends on the hydrophobic contacts involving the side chain of the conserved hydrophobic residue Phe3 and the nonpolar groups of the residues in the loop between the β-hairpin and α-helix as formed in the NMR structure. This can be demonstrated in Figure 5d which shows the percentage for the formation of the two HBs in the β-hairpin (open squares) and the percentage for the formation of the hydrophobic contacts between the side chain of the Phe3 and the residues near the loop ranging from Phe14 to Leu20 with (solid circles) and without (open circles) the constraint that the HBs is formed as a function of $N_{nl}$. One can see that the percentage for the formation of the hydrophobic contacts with the HBs formed is much higher than that without constraint, implying that the folding and stabilization of the N-terminal *β*-hairpin depend strongly on the hydrophobic contacts involving the side chain of Phe3 and the hydrophobic groups in the loop. It is interesting to note that the hydrophobic residue in the position of Phe3 is highly conserved in the classical zinc-finger motif, but the reason is still unknown. The present results suggest that stabilizing the N-terminal *β*-hairpin is one of the possible reasons for nature to select this hydrophobic residue. However, formation of such hydrophobic contacts involving Phe3 needs the formation of the loop structure ranging from Phe14 to Leu20 since this loop structure is served as the scaffold to anchor the nonpolar groups. As the packing of the hydrophobic Phe14, Leu20 and His23, which forms the hydrophobic core, is responsible for the formation and stabilization of this loop, the formations of the HBs and the hydrophobic contacts in stabilizing the N-terminal β-hairpin rely on the hydrophobic core formed by Phe14, Leu20 and His23 and the coordinations of the Cys5, Cys10 and His23 to the Zn(II). Meanwhile, the percentage for the formation of HBs and hydrophobic contacts involving Phe3 are saturated when these three coordination bonds are formed as shown in Figure 5d, indicating that the first three bonds are sufficient and necessary for the β-hairpin folding.

The effect of the formations of Zn(II)-Cys5, Zn(II)-Cys10 and Zn(II)-His23 on the folding and stability of the C-terminal α-helix is also mediated by this hydrophobic core (Figure S14d). This is easy to be understood since both the Leu20 and His23 locate at the first two helical turns of the α-helix, and the formation of the hydrophobic core, which need the formation of the first three coordination bonds, fixes these two adjacent helical turns, contributing to the folding and stabilization of the α-helix segment.

In experiments, Weiss and coworkers found that the mutation of the conserved hydrophobic residues (corresponding to the Phe14 and Leu20 of this zinc-finger) to smaller residues increases the dynamical instability of the zinc-finger dramatically[18,19]. The importance of the hydrophobic core for the folding and stabilization of the zinc-finger peptide revealed in this work is consistent with this experimental data.

Such a mechanism of hydrophobic core mediated folding and stabilization of the component secondary structures by the zinc binding ensures that one single Zn(II) can simultaneously stabilize two or more secondary structures which have large spatial separation. Although the hydrophobic core of this peptide only includes three residues (Phe14, Leu20 and His23), which is much smaller than that of the other nature proteins, this relatively weak hydrophobic interaction is largely compensated by the binding of the Zn(II), contributing to the high stability of the zinc-finger peptide. It is worth mentioning that scientists in the field of protein design have successfully designed a protein which can fold to the same structure as the target $Cys_2His_2$ zinc-finger by replacing the zinc binding site with a bigger hydrophobic core[7]. However, the designed protein has low folding cooperativity and quite different folding pathway compared with the $Cys_2His_2$ type zinc-finger[67], which again illustrates the crucial role of the zinc binding in the folding of the zinc-finger motif.

Very recently, based on quantum chemical method, Dudev and Lim calculated the relative stabilities of a number of possible coordination structures along the zinc-finger folding pathway[68]. The predicted binding order of the four native ligands is consistent with the present results. Our all-atom simulations, which focus on the coupling between the zinc binding and peptide folding, and this quantum chemical calculations are complementary and provide a full picture for the folding of the classical $Cys_2His_2$ type zinc-finger.

**Summary**

Our work provides a detailed atomistic picture for the zinc coupled folding of the zinc-finger motif in which the zinc binding directs and modulates the folding and stabilization of the component secondary and tertiary structures. The results of this work provide significant insight into the general mechanism of the metal-cofactor dependent protein folding. Such knowledge is also helpful to understand other metal induced conformational changes of biological importance. Meanwhile, the method developed in this work for studying the folding of the zinc-fingers can be used to study other metal coupled folding to which the application of the classical molecular dynamics without considering the charge transfer and Zn(II) induced potontation/deprotonation effects fails. It is also applicable to a wide range of other structure and functional problems in which the metal ion plays a major role. The knowledge about the factors that are crucial for the folding and stability of the zinc-finger revealed in this work is also helpful for simplifying the protein folding alphabet of the zinc-finger motif, which is essential for designing new zinc-fingers with certain functions[5,6,69]. In addition, the present work provides an example that the problem of the metal-cofactor dependent protein folding can be understood based on computational simulations with the complements of experimental data.

In this work, due to the computational complexity, the Zn(II) induced polarization effect is treated by an effective way. Recently, developing the polarizable force field with high accuracy and transferability is becoming the object of intensive efforts world-wide[50,70-75]. With the further improvement of the force field accuracy and simulation algorithm, as well as the development of computer speed, the simulation of protein folding with polarizable force field will be feasible. It will be interesting to implement such polarizable force fields into the present model to simulate the folding of metalloproteins in the future.

**Acknowledgements.** The authors thank Prof. M. Karplus, Prof. C. Lim and Prof. C. C. Wang for helpful discussions and suggestions. This work is supported by NSF of China (No. 10704033, 90403120, 10504012, 10474041 and 10774069), National Basic Research Program of China (2006CB910302 and 2007CB814806) and China Postdoctoral Science Foundation. Computational support is provided by Shanghai Supercomputer Center.

**Table of contents graphic:**

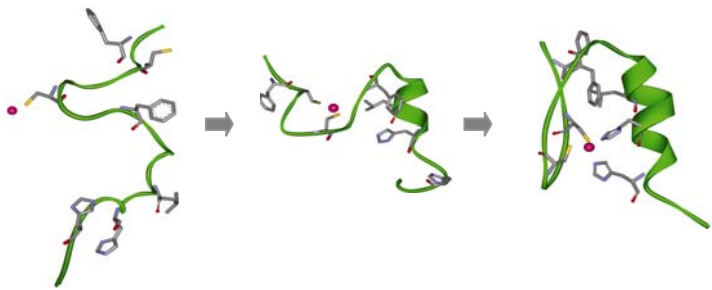